\documentclass[aps, prl,reprint,superscriptaddress,showpacs, citeautoscript,floatfix]{revtex4-1}

\usepackage{graphicx}
\usepackage{dcolumn}
\usepackage{bm}
\usepackage{natbib}
\usepackage{amsfonts}
\usepackage{amsmath}
\usepackage{amssymb}
\usepackage{xcolor}

\begin{document}

\title{Dispersion, damping, and intensity of spin excitations in the single-layer (Bi,Pb)$_{2}$(Sr,La)$_{2}$CuO$_{6+\delta}$ cuprate superconductor family}

\author{Y.~Y.~Peng}
\email{pyy2018@illinois.edu} \altaffiliation{Present address: Department
of Physics and Seitz Materials Research Laboratory, University of
Illinois, Urbana, IL 61801, USA} \affiliation{Dipartimento di Fisica,
Politecnico di Milano, Piazza Leonardo da Vinci 32, I-20133 Milano, Italy}
\author{E.~W.~Huang}
\affiliation{Department of Physics, Stanford University, Stanford,
California 94305, USA} \affiliation{Stanford Institute for Materials and
Energy Sciences, SLAC National Accelerator Laboratory and Stanford
University, Menlo Park, CA 94025, USA}
\author{R.~Fumagalli}
\affiliation{Dipartimento di Fisica, Politecnico di Milano, Piazza
Leonardo da Vinci 32, I-20133 Milano, Italy}
\author{M.~Minola}
\affiliation{Max-Planck-Institut f\"{u}r Festk\"{o}rperforschung,
Heisenbergstra$\beta$e 1, D-70569 Stuttgart, Germany}
\author{Y.~Wang}
\affiliation{Stanford Institute for Materials and
Energy Sciences, SLAC National Accelerator Laboratory and Stanford
University, Menlo Park, CA 94025, USA} \affiliation{Department of Physics, Harvard University, Cambridge 02138, USA}
\author{X.~Sun}
\affiliation{Beijing National Laboratory for Condensed Matter Physics,
Institute of Physics, Chinese Academy of Sciences, Beijing 100190, China}
\author{Y.~Ding}
\affiliation{Beijing National Laboratory for Condensed Matter Physics,
Institute of Physics, Chinese Academy of Sciences, Beijing 100190, China}
\author{K.~Kummer}
\affiliation{ESRF, The European Synchrotron, CS 40220, F-38043 Grenoble
Cedex, France}
\author{X.~J.~Zhou}
\affiliation{Beijing National Laboratory for Condensed Matter Physics,
Institute of Physics, Chinese Academy of Sciences, Beijing 100190, China}
\author{N.~B.~Brookes}
\affiliation{ESRF, The European Synchrotron, CS 40220, F-38043 Grenoble
Cedex, France}
\author{B.~Moritz}
\affiliation{Stanford Institute for Materials and Energy Sciences, SLAC
National Accelerator Laboratory and Stanford University, Menlo Park, CA
94025, USA}
\author{L.~Braicovich}
\affiliation{ESRF, The European Synchrotron, CS 40220, F-38043 Grenoble
Cedex, France}
\author{T.~P.~Devereaux}
\affiliation{Stanford Institute for Materials and Energy Sciences, SLAC
National Accelerator Laboratory and Stanford University, Menlo Park, CA
94025, USA}
\author{G.~Ghiringhelli}
\email{giacomo.ghiringhelli@polimi.it} \affiliation{Dipartimento di
Fisica, Politecnico di Milano, Piazza Leonardo da Vinci 32, I-20133
Milano, Italy} \affiliation{CNR-SPIN, Politecnico di Milano, Piazza
Leonardo da Vinci 32, I-20133 Milano, Italy}

\date{\today}

\begin{abstract}

Using Cu-$L_3$ edge resonant inelastic x-ray scattering (RIXS) we measured
the dispersion and damping of spin excitations (magnons and paramagnons)
in the high-$T_\mathrm{c}$ superconductor
(Bi,Pb)$_{2}$(Sr,La)$_{2}$CuO$_{6+\delta}$ (Bi2201), for a large doping
range across the phase diagram ($0.03\lesssim p\lesssim0.21$). Selected
measurements with full polarization analysis unambiguously demonstrate
the spin-flip character of these excitations, even in the overdoped
sample. We find that the undamped frequencies increase slightly with doping for
all accessible momenta, while the damping grows rapidly, faster in the
(0,0)$\rightarrow$(0.5,0.5) nodal direction than in the
(0,0)$\rightarrow$(0.5,0) antinodal direction. We compare the experimental
results to numerically exact determinant quantum Monte Carlo (DQMC)
calculations that provide the spin dynamical structure factor
$S(\textbf{Q},\omega)$ of the three-band Hubbard model. The theory
reproduces well the momentum and doping dependence of the dispersions and
spectral weights of magnetic excitations. These results provide compelling
evidence that paramagnons, although increasingly damped, persist across
the superconducting dome of the cuprate phase diagram; this implies that
long range antiferromagnetic correlations are quickly washed away, while
short range magnetic interactions are little affected by doping.
\end{abstract}

\maketitle

\section{Introduction}

In layered cuprates, doping charge carriers into the CuO$_2$ planes
rapidly suppresses the long-range antiferromagnetic (AF) order of the
insulating parent compounds and leads to high critical temperature
superconductivity \cite{PALeeReview}.
The proximity of antiferromagnetism to superconductivity in the phase
diagram of cuprates and other unconventional superconductors suggests the
importance and necessity of a detailed understanding of antiferromagnetism
and, more generally, spin excitations. In the absence of long range order, the most important information concerning spin excitations is encoded in their dispersion, intensity, and
broadening. Historically, inelastic neutron scattering (INS) was the
exclusive technique for studying spin order and excitations in cuprates
with momentum and energy resolution. More recently, resonant inelastic
x-ray scattering (RIXS), performed at the Cu $L_3$ resonance
\cite{AmentPRL,HaverkortPRL,LucioPRL}, has become a promising alternative that complements and extends neutron scattering results due to more favorable cross
sections and beam flux that allow for measurements on small crystals,
films, and heterostructures.

The largest intensities seen in INS studies correspond to the elastic
scattering peak at the AF-ordering wavevector
$\textbf{Q}_\mathrm{AF}$=(0.5,0.5) in undoped materials and the magnetic
resonance of the doped superconducting compounds
\cite{FongYBCO,KeimerBi2212,KeimerTl2201,GrevenHg1201}, which rapidly loses intensity in the overdoped regime \cite{WakimotoINS2004,WakimotoINS2007,Spinreview}. 
Generally, INS has demonstrated that around $\textbf{Q}_\mathrm{AF}$, both
elastic and inelastic scattering due to spin fluctuations are suppressed
by doping. Conversely, RIXS measurements have found that spin excitations
persist, upon doping, in a large momentum region around the Brillouin zone
center $\Gamma$=(0,0), even for heavily overdoped, non-superconducting,
metallic systems
\cite{TaconNP,DeanNM,DeanPRL,TaconPRB,PengPRB,MatteoPRL1}. How
can we use these results to gain insight on the possible role of spin
fluctuations in the formation of Cooper pairs needed for
superconductivity? The different trends observed in INS and RIXS can be
reconciled by noting that the two techniques primarily access different
regions of reciprocal space. As shown by numerical
calculations \cite{TomNM,HuangPRB} of the spin dynamical structure factor
$S(\textbf{Q},\omega)$, the short-range, high-energy spin excitations
measured by RIXS close to the magnetic Brillouin zone boundaries are less
relevant for pairing than the low-energy ones probed by INS around
$\textbf{Q}_\mathrm{AF}$. Altogether these findings are consistent with
arguments in favor of spin-fluctuations contributing to pairing in cuprate superconductors \cite{Scalapinoreview,DahmNP}. 

Nevertheless, a complete picture has yet to emerge, as
most results have focused primarily on the (1,0) anti-nodal direction, parallel to the Cu-O bonds. 
In contrast to that behavior, recent experiments have found that the spectra 
along the (1,1) nodal direction lack easily identifiable
collective excitations \cite{ChangLSCO2017,DeanLSCO2017}.
Interpreted as charge channel, particle-hole excitations, with support from random-phase approximation (RPA) calculations, these findings challenge the existence of paramagnons (damped collective spin
excitations) in doped cuprates \cite{GrioniNM,DeanBi2014,ChangLSCO2016,ChangLSCO2017,DeanLSCO2017}.
The apparent dichotomy between dispersing spin excitations along (1,0), largely insensitive to doping, and a continuum of charge modes along (1,1), which apparently soften upon doping \cite{GrioniNM,DeanBi2014}, raises questions concerning the doping evolution of magnons into paramagnons and the correct microscopic description of spin excitations in overdoped compounds.
The dichotomy may be even more interesting in light of our recent RIXS study demonstrating a direct correlation across several cuprate families between optimal $T_{c}$ and the difference in dispersion along the (1,0) and (1,1) directions in AF parent compounds \cite{AFcuprate}, where the difference also correlates with parameters of microscopic models that involve the oxygen degree of freedom, {\it e.g.} the charge-transfer energy \cite{Wang2018}.

While one can debate whether or not a given microscopic model properly describes the low energy physics in the cuprates, these models do provide useful insight on the impact of correlations and their evolution with doping.  In particular, recent results from determinant quantum Monte Carlo (DQMC) simulations of the single-band Hubbard model challenge conventional wisdom about that rate at which the influence from correlations decreases with carrier doping \cite{Kung2015}.  Based solely on single-particle properties \cite{Kohno,WangPRB2018}, one would conclude that correlations weaken rapidly with doping, revealing Fermi-liquid-like behavior just into the overdoped regime.  However, the behavior of multi-particle spin and charge response functions shows that the influence from correlations can persist to relatively high doping levels.  When compared with the results from RPA calculations, the influence of correlations persists across the Brillouin zone and throughout the doping range relevant to the cuprates.  Additional model calculations compared to RIXS experimental results demonstrate a clear delineation between the low energy spin and charge excitations \cite{Huang2016}.  Given clear distinctions from model calculations, and another recent proposal for a low-energy spin excitation in overdoped cuprates from RPA calculations, which may be resolvable with an improved energy resolution ($\sim$60 meV) \cite{ChangLSCO2016}, an extensive high-resolution study of the doping and momentum dependence of paramagnons may reconcile these differing perspectives.

In this article we present a systematic RIXS study of magnetic excitations
in single-layer (Bi,Pb)$_{2}$(Sr,La)$_{2}$CuO$_{6+\delta}$ (Bi2201), with 4 doping levels ranging from the AF insulator
to the overdoped superconductor. Our data cover a significant portion of
reciprocal space with an energy resolution of about 55 meV. Polarization-resolved measurements demonstrate the spin-flip nature, even in the
overdoped region, of the main spectral feature commonly assigned to
paramagnons. We extract the paramagnon dispersion, damping, and spectral
weight as functions of momentum and doping by fitting the spectra with a
general function valid for all damping regimes \cite{LamsalPRB,
ChangLSCO2017}. We find that both the undamped frequency and damping
factor increase with doping. Moreover, the damping and the spectral weight
display a significant momentum dependence. These observations are captured
by DQMC calculations of the spin
dynamical structure factor $S(\textbf{Q},\omega)$ for the three-band
Hubbard model, which allow us to discuss quantitatively the implications
of the experimental results.

\begin{figure}[tbp]
\begin{center}
\includegraphics[width=0.8\columnwidth,angle=0]{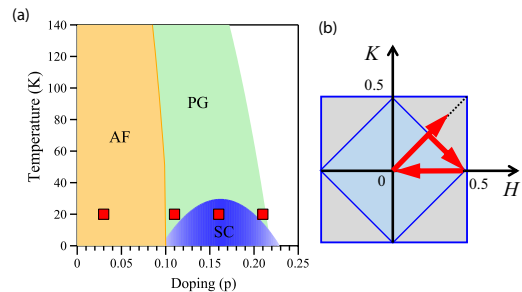}
\end{center}
\caption{(a) Schematic temperature-doping phase diagram of (Bi,Pb)$_2$(Sr,La)$_2$CuO$_{6+\delta}$. It shows the antiferromagnetic (AF), superconducting (SC) and the pseudogap (PG) regions. Here we study four doping levels as indicated by the solid red squares. (b) 2D reciprocal lattice for the pseudotetragonal structure and the first Brillouin zones (structural in light grey, magnetic in light blue). Coordinates $H$ and $K$ are in r.l.u.. The path followed for the measurements is indicated by the red arrows, starting at (0.25,0.25) and ending around (0.30,0.30) via (0.5,0) and (0,0).
\label{Fig1}}
\end{figure}

\section{RIXS experiment}

\subsection{Experimental details}

We have studied four doping levels of
Bi2201 as indicated in the phase diagram \cite{NMRBi2201} in Fig.~1(a):
antiferromagnetic (AF, \emph{p}$\simeq0.03$), underdoped with $T_c$=15K (UD15K, \emph{p}$\simeq0.11$), 
optimal doping with $T_c$=33K (OP33K, \emph{p}$\simeq0.16$) and 
overdoped with $T_c$=11K (OD11K, \emph{p}$\simeq0.21$). The sample growth and
characterization methods have been reported previously
\cite{PengNaturecomm,meng,zhaoCPL}. The RIXS measurements were performed
with the ERIXS spectrometer at the beam line ID32 of the European
Synchrotron Radiation Facility (ESRF) in Grenoble, France \cite{NickERIXS}. The RIXS spectra were collected at 20K
with $\pi$ incident polarization (parallel to the scattering plane) to
maximize the single-magnon signal \cite{AmentPRL,LucioPRB}. The scattering
angle was fixed at $149.5^\circ$ and the incident photon energy was tuned
to the maximum of the Cu $L_3$ absorption peak around 931 eV. The total
experimental energy resolution was about 55 meV. The samples were cleaved
in air a few minutes before installation in the measurement vacuum
chamber. The reciprocal lattice units (rlu) used in figures and in text
below are defined using the pseudotetragonal unit cell with
$a=b=3.86$~{\AA}. The zero energy-loss position of each spectrum was
determined by measuring, with the same incident photon energy, one
nonresonant spectrum of silver paint or carbon tape.

\subsection{RIXS data overview and fitting procedure}

\begin{figure}[tbp]
\begin{center}
\includegraphics[width=0.8\columnwidth,angle=0]{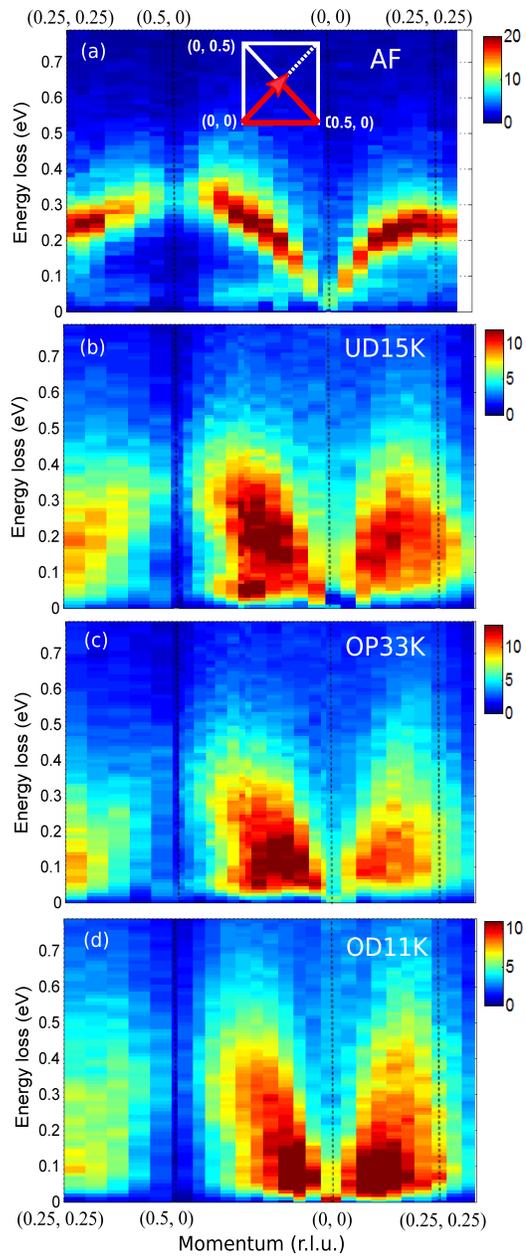}
\end{center}
\caption{Energy/momentum intensity maps of RIXS spectra for (a) AF
 (\emph{p}$\simeq$0.03), (b) UD15K (\emph{p}$\simeq$0.11), (c) OP33K
 (\emph{p}$\simeq$0.16), and (d) OD11K (\emph{p}$\simeq$0.21) along the
 high-symmetry momentum trajectory indicated in Fig.\ref{Fig1}b and in the inset of (a). The
 intensity is in unit of photons/s/eV. Data were taken with
 $\pi$-polarized incident light at 20 K. Elastic peaks were subtracted for
 a better visualization of the low energy features.
\label{Fig2}}
\end{figure}

Figure~\ref{Fig2} displays the energy/momentum intensity maps of RIXS
spectra for AF (a), UD15K (b), OP33K (c) and OD11K (d) along the
high-symmetry directions indicated in the inset of panel (a). The magnetic
excitations are very sharp for the AF case and become increasingly broader
with doping, in agreement with previous results
\cite{TaconNP,DeanNM,DeanLSCO2017}. The strongly suppressed intensity near
\textbf{Q}=(0.5,0) is consistent with the anomalous broadening and damping
of spin waves at that momentum in square 2D AF lattice observed previously
in, e.g., La$_2$CuO$_4$ and in copper deuteroformate tetradeurate by INS
\cite{INSLCO,Christensen2007} and in CaCuO$_2$ with RIXS, which
was ascribed to the decay of spin waves into fractional spin excitations
\cite{DallaPiazza2015}. We note that we do not observe any of the
low-energy spin excitation along (0,0)$\rightarrow$(0.5,0.5) predicted by
the RPA calculations \cite{ChangLSCO2016}. On the other hand, we have
observed a sharp charge-order peak along (0,0)$\rightarrow$(0.5,0)
direction in the overdoped Bi2201 as reported in
Ref.~[\onlinecite{COinOD}]. In the present work we have
subtracted out the elastic peak to focus on the study of magnetic
excitations.

\begin{figure}[htbp]
\includegraphics[width=\columnwidth,angle=0]{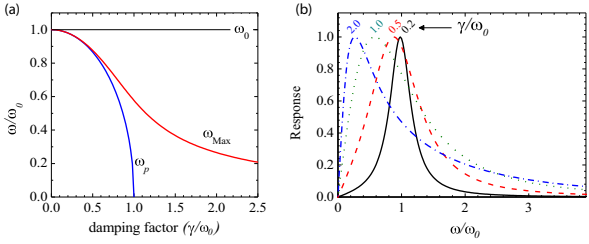}
\caption{(a) Illustration of the relative deviation made, for a generic damped oscillator, when using the peak position $\omega_{\mathrm{Max}}$ or the central frequency of an antisymmetrized Lorentzian $\omega_p$ instead of the actual undamped frequency $\omega_0$, as function of the damping factor $\gamma/\omega_0$. (b) The effect of damping on the shape of the response of the damped oscillator: the underdamped ($\gamma/\omega_0=0.2$, black curve) symmetric peak at $\omega = \omega_0$ moves towards lower frequency for increasing $\gamma$. For $\gamma/\omega_0>1$ it cannot be fitted anymore by an antisymmetrized Lotentzian with poles at $\omega=\pm \omega_p$ and the spectral shape becomes highly asymmetrical (blue curve, $\gamma/\omega_0=2.0$).
\label{Fig3}}
\end{figure}

For a quantitative analysis of these experimental data we have used a
general fitting procedure applicable to all cases for the extraction of
the energy, intensity and broadening of the spin excitations. The RIXS
process leading to a spin excitation can be expressed in terms of the
magnetic susceptibility $\chi$, in strict analogy with INS experiments: the spin dynamic structure factor
$S(\textbf{Q},\omega)$ determines the scattering cross section,
which is proportional to the imaginary part of the susceptibility $\chi''
(\textbf{Q},\omega)$. The microscopic scattering process is very different
for RIXS and INS, so that absolute intensity cannot be directly
compared. However their relative intensity can be compared because their
dependence on the scattering angles and polarization of the scattering
particles are known to evolve slowly with \textbf{Q} within a given
Brillouin zone \cite{AmentPRL}. Therefore we can fit the RIXS spectra to
obtain relevant estimates of the energy, width, and relative intensity of
the spin-flip peak and, ultimately, of $\chi$. The fitting function is
easily obtained from the expressions of $\chi$ and of its imaginary part
$\chi''$. For a generic damped harmonic oscillator, of given undamped
frequency $\omega_0$ and damping factor $\gamma$, it is well known that
the complex susceptibility is  $\chi(\omega) \propto 1 / [(\omega_0 ^2 -
\omega ^2) + 2 i \gamma \omega]$. For a given \textbf{Q} we can thus write

 \begin{equation}\label{Eq1}
   \chi'' (\textbf{Q},\omega) \propto \frac{\gamma \omega}{(\omega ^2 - \omega_0 ^2)^2 + 4 \gamma ^2 \omega ^2 }
 \end{equation}

When the damping is not too large (i.e. underdamped, $\gamma < \omega_0$) the shape of
$\chi'' $ can be reproduced by an antisymmetrized Lorentzian function
$L(\omega)$, i.e., the difference of two Lorentzian peaks at position $
\pm\omega_p$ and same width $\gamma$:

\begin{equation}\label{Eq2}
  L(\omega) = \frac{\gamma }{(\omega - \omega_p)^2 + \gamma ^2 } - \frac{\gamma }{(\omega + \omega_p)^2 + \gamma ^2 }
\end{equation}

Indeed the two functions are identical up to a normalization factor if $\omega_p^2 = \omega_0^2 -
\gamma^2$, which is possible only for $\gamma \leq \omega_0$. As pointed
out by  Lamsal et al. \cite{LamsalPRB}, in some recent RIXS papers
\cite{TaconNP,DeanNM} the antisymmetrized Lorentzian function has been
used to fit damped paramagnon curves, leading to an inaccurate estimation of $\omega_0$. The deviation is evident in the case of overdamped
paramagnons (i.e. $\gamma > \omega_0$), which cannot be fitted by $L(\omega)$ in a satisfactory way. But even for underdamped paramagnons, that can be fitted well by the
antisymmetrized Lorentzian, a non-negligible deviation is made if one assigns
to $\omega_0$ the value $\omega_p$ obtained from the fitting. It must be
also noted that $\omega_\mathrm{Max}$, the maximum of the function
$\chi''$ of Eq.~(\ref{Eq1}), is different from both $\omega_0$ and
$\omega_p$ when $\gamma \sim \omega_0$ (critically damped), and thus it cannot be used to
evaluate ``by eye'' the undamped frequency either. In Fig~\ref{Fig3} we
present the relative deviation of $\omega_p$ and $\omega_\mathrm{Max}$ as
function of the damping factor $\gamma/\omega_0$. Therefore we have
consistently fitted all our paramagnon spectra with the function $\chi''$
in Eq. (\ref{Eq1}), convoluted with the experimental resolution function, obtaining the values of $\omega_0$, $\gamma$ and
relative intensity presented and discussed below.

\begin{figure}[tbp]
\begin{center}
\includegraphics[width=\columnwidth,angle=0]{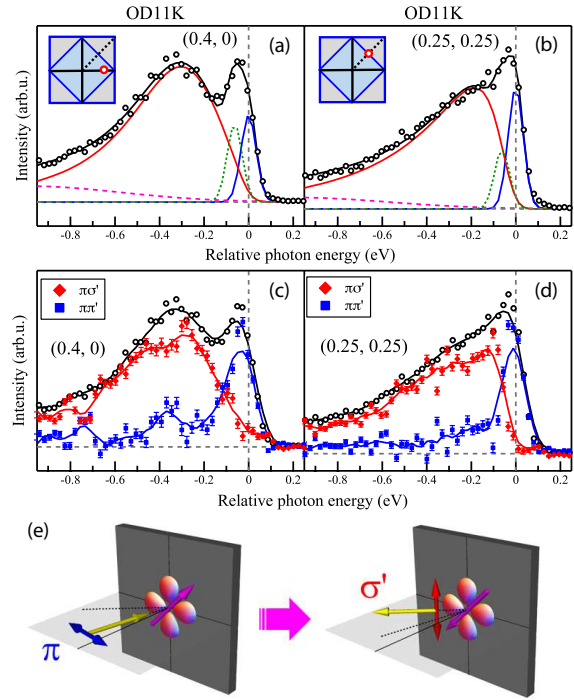}
\end{center}
\caption{(a,b) RIXS spectra at (0.4, 0) and (0.25,0.25) respectively, indicated by the red circles in the insets, measured with $\pi$-polarized incident light at 20 K for OD11K ($p\simeq0.21$). The spectra are decomposed into the magnetic excitation (red line), the elastic scattering (blue line), the phonon scattering (green dotted line), and the charge background (dashed magenta line). (c,d) Polarization resolved measurements for OD11K ($p\simeq0.23$) with incident $\pi$-polarized light. Statistical error bars are calculated from the number of photon counts. (e) Schematic illustration of the spin-flip process: the angular momentum conservation requires the $90^\circ$ rotation of the photon polarization, which has maximum intensity in the $\pi \sigma\prime$ channel at positive momenta (close to normal incidence, grazing emission). The spin conserving processes can be found only in the $\pi \pi\prime$ channel. Here $\sigma\prime$ and $\pi\prime$ refer to the scattered x-ray polarization.
\label{Fig4}}
\end{figure}

\section{RIXS spectra and fitting results}

\begin{figure*}[htbp]
\begin{center}
\includegraphics[width=2\columnwidth,angle=0]{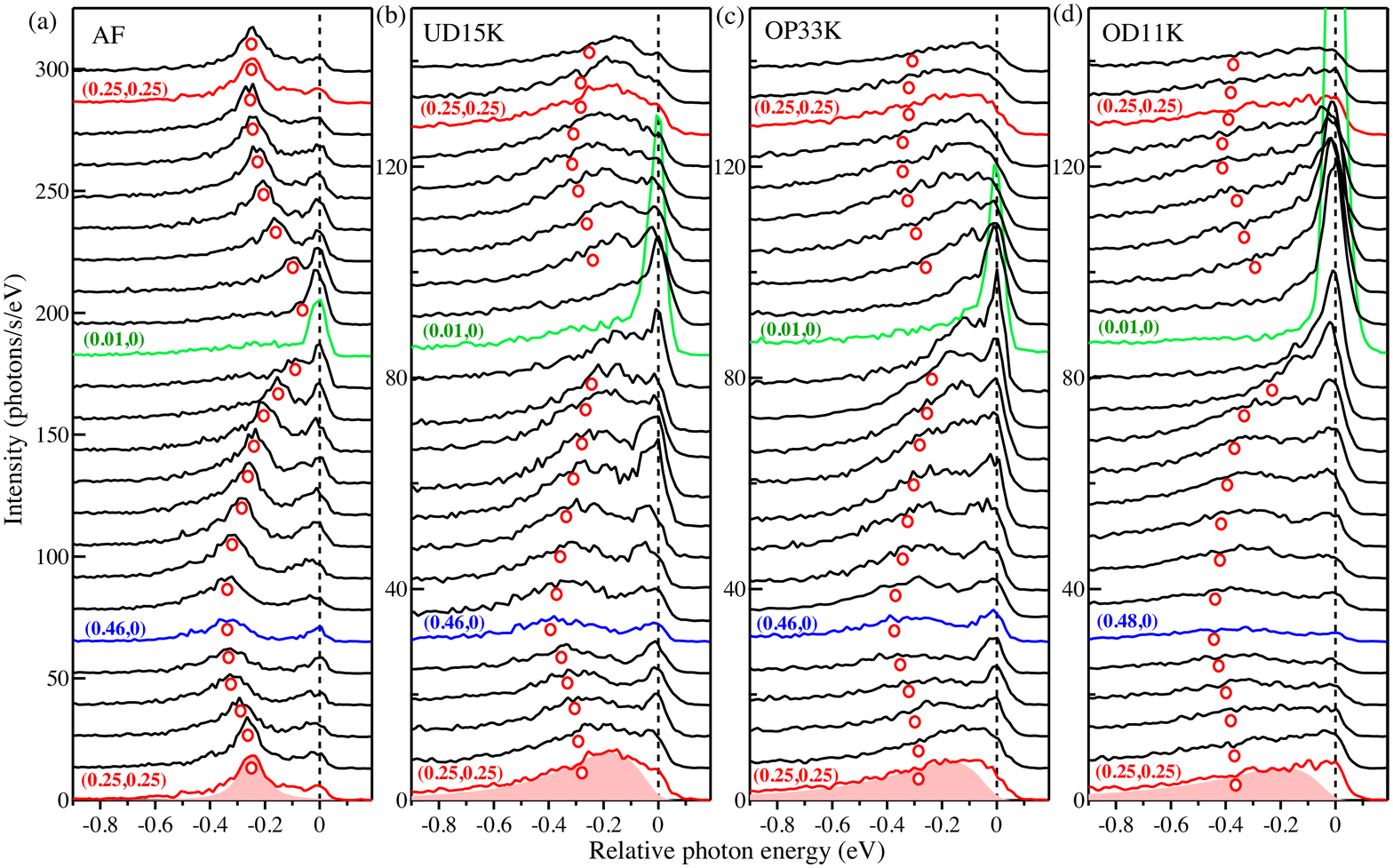}
\end{center}
\caption{ The raw RIXS spectra for AF (a), UD15K (b), OP33K (c) and OD11K
(d) along the high-symmetry directions indicated in Fig.~\ref{Fig1}b. Each
spectrum is shifted vertically for clarity. Circles denote the undamped
frequency of magnetic excitations determined from fittings. The red shaded
areas in the bottom spectra represent the magnetic excitation.
\label{Fig5}}
\end{figure*}

Figure \ref{Fig4}(a,b) show selected examples of fitting for OD11K at the two
representative momenta \textbf{Q}=(0.4,0) and \textbf{Q}=(0.25,0.25). As
expected the paramagnon excitation (red line) dominates the mid-infrared
range. To validate the assignment of the fitted intensity to spin
excitations we exploit the polarimeter of the ERIXS spectrometer
\cite{LucioPolarimeter}. In fact, the spin-flip scattering is accompanied
by a $90^\circ$ rotation of the photon polarization as shown in
Fig.~\ref{Fig4}(e). The polarimeter spectra in Fig.~\ref{Fig4}(c,d)
demonstrate that the crossed polarization channel $\pi\sigma'$ (with $\sigma'$
refers to the scattered x-ray polarization) dominates
the mid-infrared region, even in the absence of a well-defined peak as in
\textbf{Q}=(0.25,0.25), confirming that the spectra are strongly dominated
by spin flip excitations. 
On the other hand, the quasi-elastic peak is
spin-conserving (blue lines for the $\pi \pi'$ scattering), and a non-negligible non-spin-flip intensity is present in the mid-IR region too,
due to the charge continuum and to bi-paramagnons.

\begin{figure}[htbp]
\begin{center}
\includegraphics[width=\columnwidth,angle=0]{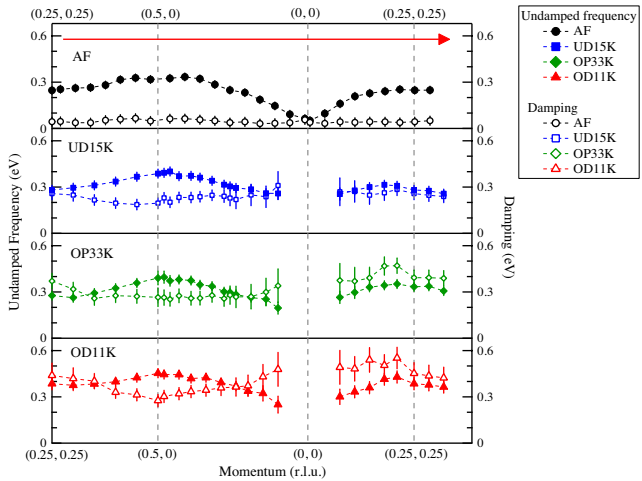}
\end{center}
\caption{Doping dependence of undamped frequency $\omega_0$ (solid symbols) and
damping $\gamma$ (hollow symbols) for magnetic excitations along the high-symmetry
directions indicated in Fig.~\ref{Fig1}b. The error bars represent
the statistical error from the fitting procedure.
\label{Fig6}}
\end{figure}

The raw RIXS spectra for the four dopings are shown in Fig.~\ref{Fig5}.
The magnetic excitations change dramatically: the sharp peaks of the AF
sample become broader in UD15K, show loose peak profiles in OP33K, and
eventually change to long tails in OD11K. This can be seen most clearly in
the bottom spectrum at \textbf{Q}=(0.25,0.25): the paramagnon (red
shading) changes from a peak in AF to a heavily damped mode in OD11K. We
have fitted all the spectra with the procedure explained in Set. II. The red
circles in the spectra indicate the undamped frequency given by the
fitting. The undamped freqency $\omega_0$ and the damping rate $\gamma$ for all spectra
are summarized in Fig.~\ref{Fig6}. We do not report the fitting results
for the spectra very close to (0,0), where the uncertainty is too large due to the elastic peak.
The evolution of magnetic excitations with doping and momentum can thus be
assessed quantitatively: the dispersion is well defined in AF with small
damping rates along the whole trajectory in reciprocal space; in UD15K, the damping increases
significantly and becomes comparable with the undamped frequency in the
(0,0)$\rightarrow$(0.5,0.5) direction; upon further doping, the damping
becomes larger than the undamped frequency for OP33K and OD11K along the
nodal direction, where paramagnons become overdamped spin-flip modes as
demonstrated above with the polarimeter. This is consistent with the recent RIXS study
showing how spin excitations in cuprates evolve from collective paramagnons to
incoherent spin-flip excitations across optimal doping
\cite{MatteoPRL2}.  Along the AF Brillouin zone
direction it is noteworthy that the crossing point between the undamped
frequency and the damping moves away from (0.25,0.25) towards (0.5,0) with
increasing doping, indicating that the overdamped region expands with
doping from the nodal direction, possibly from the AF point (0.5,0.5). The
damping increase upon doping seems stronger along the nodal than along the
antinodal direction. This fact most likely comes from the increase of the
scattering of spin excitations with the electron-hole continuum
\cite{Ramantheory}, as well as from the contribution of incoherent
particle-hole excitations to the RIXS spectra \cite{MatteoPRL2}, which
might be anisotropic in cuprates.

\section{Determinant quantum Monte Carlo calculation}

\begin{figure}[tbp]
\begin{center}
\includegraphics[width=\columnwidth,angle=0]{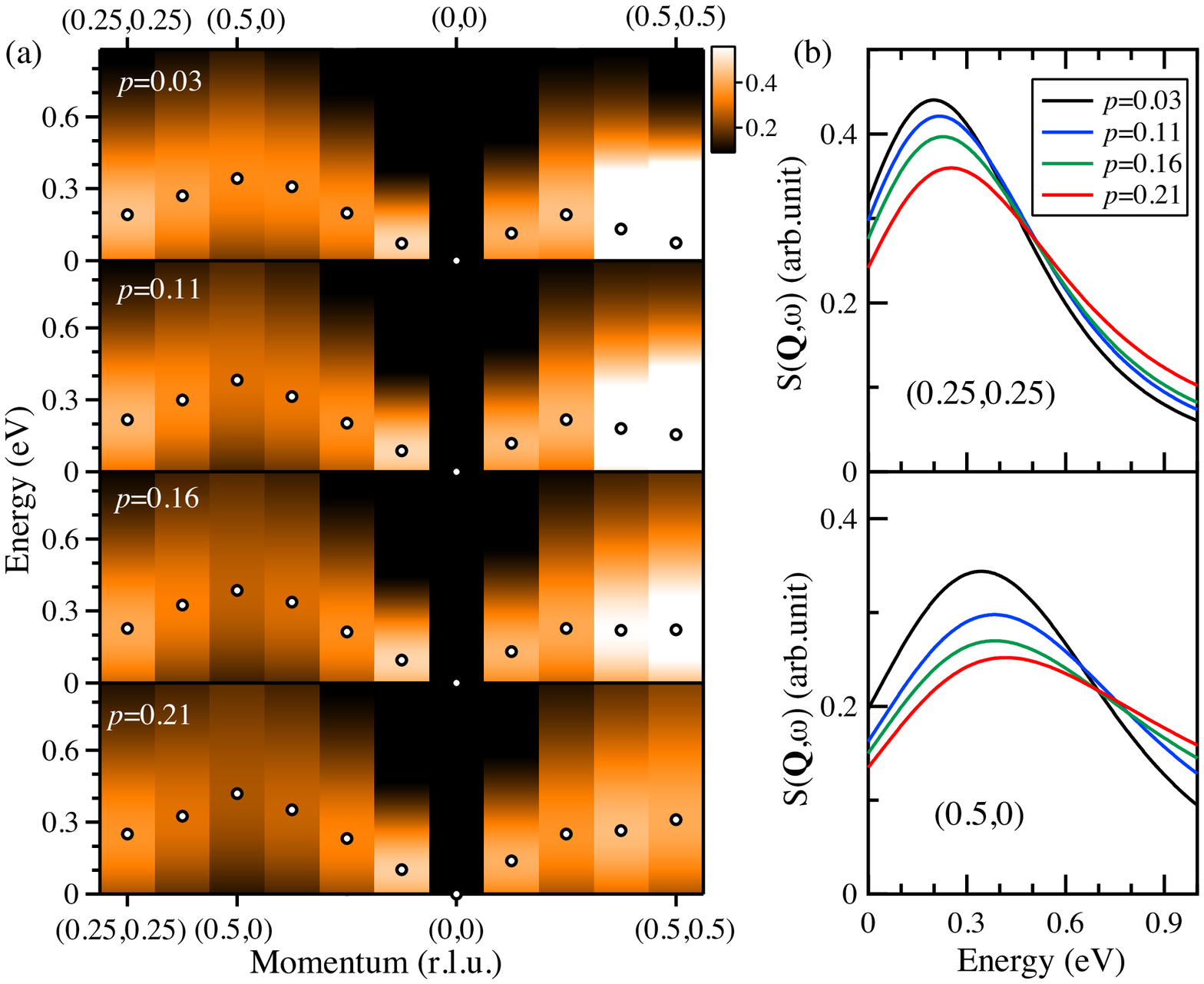}
\end{center}
\caption{The spin dynamical structure factor $S(\textbf{Q},\omega)$
calculated using DQMC for the three-band Hubbard model. (a) False colour
plots of the spectra along high-symmetry directions for four dopings
\emph{p}$\simeq$0.03, 0.11, 0.16 and 0.21, respectively. Black circles
indicate the peak positions. (b) The $S(\textbf{Q},\omega)$ at two high
symmetry momenta \textbf{Q}=(0.25,0.25) (top panel) and
\textbf{Q}=(0.5,0) (bottom panel), showing its evolution with doping. The
results were obtained with the three-band Hubbard model (
$U_{d}$=10.2 eV, $U_{p}$=5.9 eV, $t_{pd}$=1.35 eV, $t_{pp}$=0.59 eV,
$\Delta$=3.9 eV, \emph{T}= 0.15 eV).
\label{Fig7}}
\end{figure}

Here we employ the numerically exact DQMC method \cite{DQMCcalculation,DQMCWhite,DQMC3band,DQMCstripes}
to study the momentum and doping dependence of
$S(\textbf{Q},\omega)$ for the three-band Hubbard model with a typical set
of parameters as given in the caption of Fig.~\ref{Fig7}. Maximum entropy analytic
continuation \cite{Maxentropyref} is used to extract
$S(\textbf{Q},\omega)$ from the imaginary time correlators measured in
DQMC. The DQMC calculations show magnetic excitations that persist with
doping from \emph{p}=0.03 to \emph{p}=0.21 (Fig.~\ref{Fig7}(a)).
Fig.~\ref{Fig7}(b) shows the spectra at two representative momenta,
\textbf{Q}=(0.25,0.25) and \textbf{Q}=(0.5,0). The spectral weight of
$S(\textbf{Q},\omega)$ decreases and shifts to higher energy with doping.
The broad width of spectra is set predominantly by the high temperature in
the simulation. The three-band DQMC calculations correctly reproduce the
higher energy of magnetic excitation at \textbf{Q}=(0.5,0) relative to
\textbf{Q}=(0.25,0.25), whereas one-band calculations give nearly the same
energy at both momenta due to the more Heisenberg-like
physics \cite{TomNM}.

\section{Discussion}

\begin{figure}[tbp]
\begin{center}
\includegraphics[width=\columnwidth,angle=0]{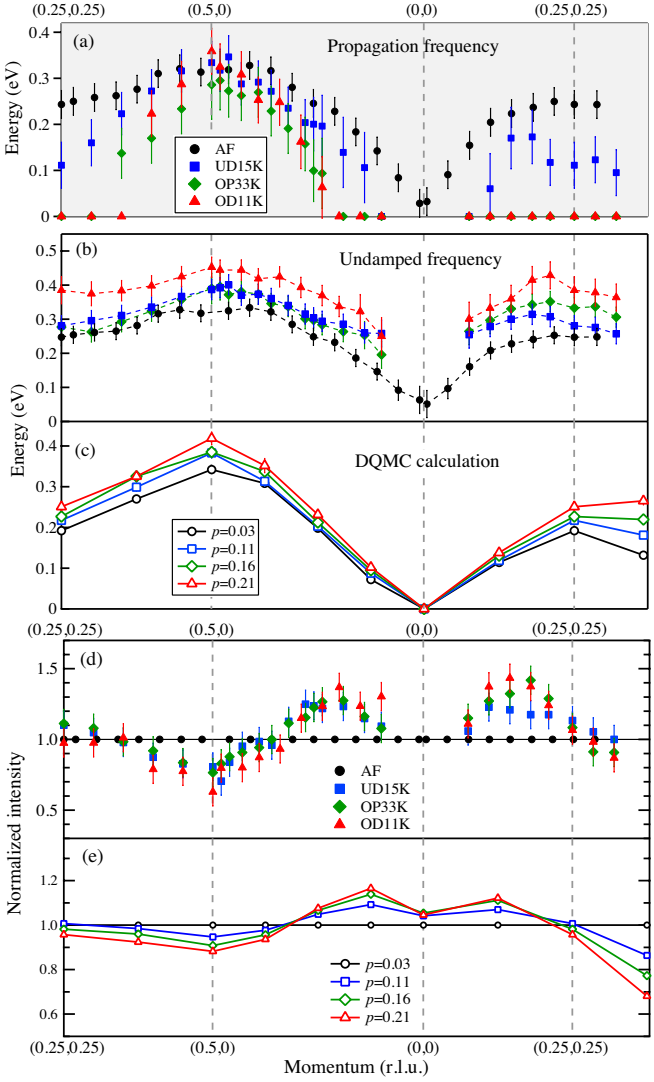}
\end{center}
\caption{Comparison of experimental and theoretical results for the spin excitations at 4 doping levels along high symmetry directions. (a) Propagation frequency $\omega_p$ and (b) undamped frequency  $\omega_0$ from fittings, and (c) peak position of the computed $S(\textbf{Q},\omega)$. (d) Experimental and (e) calculated intensities normalized, at each momentum, to the magnon intensity for the AF sample / undoped calculated case. The normalization corrects for the momentum dependent self-absorption effects in the experiment and highlights that hole doping increases short-range spin correlation and destroys the long-range one.
 \label{Fig8}}
\end{figure}

In Figure~\ref{Fig8} we compare the experimental and theoretical results.
The undamped frequency $\omega_0$ shown in panel (b) is in good
agreement with the evolution of the peak position of
$S(\textbf{Q},\omega)$ reported in panel (c): $\omega_0$ increases with
doping for all $\textbf{Q}$ values, in qualitative accord with the DQMC
calculations. The agreement is less good only in the neighborhood of
(0,0), where the experimental data are more difficult to analyze. On the
contrary along the antiferromagnetic zone boundary (AFZB) both experiment and theory find that the
dispersion is unaffected by doping, with a rigid shift of the curves to
higher energies in the $(0.25,0.25)\rightarrow(0.5,0)$ path, so that the
energy difference $\Delta E$=$\omega_0$(0.5,0)-$\omega_0$(0.25,0.25) is almost constant with
doping. This is in distinct contrast to the propagation frequency $\omega_p$ shown in
Fig.~\ref{Fig8}(a). Along the $(0,0) \rightarrow (0.5,0)$ direction, the
propagation frequency decreases slightly with doping, showing a softening
behavior as in prior results \cite{TaconNP,DeanNM}; on the other hand, the
propagation frequency along the $(0,0) \rightarrow (0.5,0.5)$ direction
decreases significantly in UD15K and goes to zero in OP33K and OD11K, as
reported for overdoped La$_{1.77}$Sr$_{0.23}$CuO$_{4}$
\cite{ChangLSCO2016}. Along the AFZB direction, the propagation
energy difference in UD15K increases by $\sim 0.1$ eV with respect to the
AF case. Notably, this is similar to the report of a larger zone-boundary
dispersion in underdoped La$_{1.88}$Sr$_{0.12}$CuO$_{4}$ than in the
parent compound La$_2$CuO$_4$ \cite{ChangLSCO2017}. It appears
evident that the propagation frequency collapses to zero in the nodal
direction when reaching the optimal doping,
whereas the undamped frequency and, more importantly, the damping grow
with doping, thus drastically changing the spectral shape of spin
excitations. Therefore the short-range magnetic interaction is little
affected by hole doping, but the collective spin excitations (paramagnons)
become increasingly damped eventually losing their propagating character.

\begin{figure}[tbp]
\begin{center}
\includegraphics[width=0.8\columnwidth,angle=0]{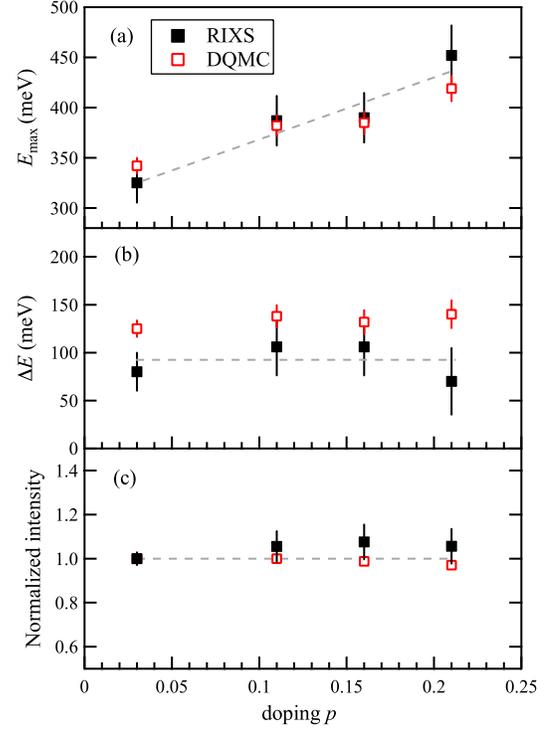}
\end{center}
\caption{Doping dependence of (a) the zone boundary undamped frequency
$E_\mathrm{max}$=$\omega_0$(0.5,0), (b) the energy dispersion along the AF zone
boundary $\Delta E$=$\omega_0$(0.5,0)-$\omega_0$(0.25,0.25), and (c) the sum of the relative
intensities of Fig.~\ref{Fig8}(d,e). The solid squares are from experiments
and hollow squares are from DQMC calculations. Dashed lines are guides for
the eye.\label{Fig9}}
\end{figure}

In Figs. \ref{Fig8}(d,e) we compare, for experiment and theory, the
intensity variations, with respect to the AF case, upon doping.  We
normalized the spectral weights to that of the AF cases to avoid possible
spurious effects in the measured data, such as self-absorption. The
agreement between experimental and numerical trends is remarkably good. The
decrease of intensity with doping when approaching
$\textbf{Q}_\mathrm{AF}$ is due to the disappearance of antiferromagnetic
correlation. It is clearly visible in the calculations and known from INS
measurements, and is hinted in the RIXS data though they cannot reach
$\textbf{Q}_\mathrm{AF}$. It has been proposed that the decrease with doping of the
spectral weight around $\textbf{Q}_\mathrm{AF}$ leads to the reduction of
the \emph{d}-wave spin-fluctuation pairing strength \cite{HuangPRB}. More surprisingly,
we find that the intensity decreases at (0.5,0) and increases around
(0,0), probably due to a strengthening of the ferromagnetic correlation
that would peak at $\textbf{Q}=0$. Remarkably, the resulting crossing
points coincide in experiment and theory around $(0.35,0)$ and
$(0.25,0.25)$. 

Figure \ref{Fig9} gives an overview of the doping dependence of magnetic
excitation properties, showing a very good agreement between RIXS
experiments and DQMC calculations. The undamped frequency at the boundary
$E_\mathrm{max}$=$\omega_0$(0.5,0) increases with doping, while the energy
dispersion along the AFZB, i.e. $\Delta E$, remains substantially
unchanged with doping. The hardening of $E_\mathrm{max}$ with doping
is effected by the three-site exchange term, which increases the overall
energy cost of spin excitations to break both spin exchange and three-site
bonds \cite{TomNM}. Although a discrepancy in the absolute value of
$\Delta E$ between calculation and experiment is still present and might
be reduced by tuning the parameters used in the simulation, the overall
constant trend vs. doping is already very similar. This result can be
explained by noting that $\Delta E$ is determined by the bare parameters
related to the charge-transfer energy, which is not significantly modified with
doping. Our recent RIXS study on undoped cuprates demonstrated $\Delta E$ was
positively correlated with the range of in-plane exchange couplings
\cite{AFcuprate}. The marginal changes in $\Delta E$ upon doping imply
that the exchange-coupling ranges are encoded in the parent compounds.

On the other hand, the energy- and momentum-integrated intensity of
paramagnons in the range accessible to RIXS is found to be constant with
doping, even though the spectral weight at $\textbf{Q}_\mathrm{AF}$ drops
due to the falloff of antiferromagnetism. This indicates that spectral
weight away from $\textbf{Q}_\mathrm{AF}$ redistributes upon doping while
roughly maintaining a constant total weight. While this is generally in
line with previous RIXS results on the persistence of spin excitations
upon doping \cite{TaconNP,DeanNM,DeanLSCO2017}, this observation is not
immediately obvious from any sum-rule-type or similar analysis. On the
contrary, INS data at $\textbf{Q}_\mathrm{AF}$ would rather suggest a
general decrease of the spin spectral weight.

\section{Conclusions}

In summary, we have revealed that the short and mid range exchange interaction is relatively little
affected by doping. This can be seen by the flatness of $\Delta E$ and by
the increase of the maximum of the paramagnon energy at (0.5,0), largely
due to the contribution from the three-site exchange that overcompensate
for the decrease of effective magnetic neighbouring sites. The melting of
the long range AF correlation is encoded in the sharp increase of the
damping. Spin excitations get increasingly coupled to charge modes and
cannot propagate more than few lattice units in plane, although their
exceptionally high energy is fully preserved even in the overdoped
samples. The three-band DQMC calculation reproduces qualitatively the
paramagnon dispersions and intensity dependence with doping and momentum.
By studying the three-band Hubbard model, and not making a priori
assumptions about the importance of various spin-exchange processes, we
have a thorough microscopic description of the electronic degrees of
freedom in the CuO$_2$ layers, with calculations capturing the same
electronic effects and processes revealed by the RIXS experiment.

It is interesting to make a connection between the variation of $T_c$ with doping and the overall evolution of the spin fluctuation spectra measured here. In a weak coupling picture in which the spin fluctuation spectrum is treated as the pairing boson in analogy with phonons in conventional superconductors, the redistribution of spectral weight shown in Figure 8 would have a strong effect on the $d$-wave pairing interaction. Since spin fluctuations carrying momenta $\textbf{Q}_\mathrm{AF}$ contribute largest to the pairing interaction, and those  with $\textbf{Q} \sim 0$ give a negative contribution to pairing, one can infer from our results that the overall strength of pairing decreases with doping as spectral weight transfers towards ferromagnetic correlations over anti-ferromagnetic ones. This is irrespective of the effect of damping of paramagnons, which only redistributes spectral weight in energy and is a subdominant effect compared to the momentum-dependent spectral weight transfer. However we caution that this conclusion can only be speculative. For example it is known from many numerical studies that various candidate ground states having different orders - in the form of stripes, charge/spin order, as well as superconductivity all exist at relatively the same energy, and therefore a full understanding of superconductivity would not be captured from simply an examination of the spin fluctuation spectra alone. Indeed recent calculations suggest an intimate coupling between charge density waves (stripes) and superconducting order in the single-band Hubbard model \cite{HubbardSC}. It would be quite useful to likewise perform an analysis of the charge degrees of freedom to further investigate a connection to superconductivity. This remains a topic for future study.

\section{ACKNOWLEDGMENTS}

\begin{acknowledgments}
We acknowledge insightful discussions with Matthieu Le Tacon, Mark Dean,
Krzysztof Wohlfeld and Jos\'e Lorenzana. The experimental data were collected at the beam
line ID32 of the European Synchrotron (ESRF) in Grenoble (F) using the
ERIXS spectrometer designed jointly by the ESRF and Politecnico di Milano.
This work was supported by ERC-P-ReXS project (2016-0790) of the
Fondazione CARIPLO and Regione Lombardia, in Italy. M. M. was partially
supported by the Alexander von Humboldt Foundation. XJZ thanks financial
support from the National Natural Science Foundation of China (11334010
and 11534007), the National Key Research and Development Program of China
(2016YFA0300300) and the Strategic Priority Research Program (B) of
Chinese Academy of Sciences (XDB07020300). EWH, YW, BM, and TPD were
supported by the U.S.~Department of Energy (DOE), Office of Basic Energy
Sciences, Division of Materials Sciences and Engineering, under Contract
No.~DE-AC02-76SF00515. Computational work was performed on the Sherlock
cluster at Stanford University and on resources of the National Energy
Research Scientific Computing Center, supported by the U.S.~DOE under
Contract No.~DE-AC02-05CH11231.
\end{acknowledgments}

\end{document}